\documentclass[sigconf,nonacm]{acmart}

\usepackage{pvldb}%
\renewcommand\vldbdoi{XX.XX/XXX.XX}
\renewcommand\vldbpages{XXX-XXX}
\renewcommand\vldbavailabilityurl{https://github.com/Delilah1230/octopus_vision} 

\AtBeginDocument{%
  }

\usepackage{amsmath}
\usepackage{graphicx} 
\usepackage{tabularx}
\usepackage{booktabs,multirow,array}
\newcolumntype{L}[1]{>{\raggedright\arraybackslash\hspace{0pt}}p{#1}}
\newcolumntype{Y}[1]{>{\raggedright\arraybackslash\hsize=#1\hsize}X}\usepackage[normalem]{ulem} 

\usepackage{listings}
\usepackage{xspace}
\usepackage[nointegrals]{wasysym}
\usepackage{xcolor}
\usepackage{enumitem}
\usepackage{tikz}
\usetikzlibrary{arrows.meta, shapes.symbols}

\newcommand{\system}{\textsc{OCTOPUS}\xspace}

\newcommand{\newstuff}[1]{\textcolor{red}{#1}}

\begin{document}
\title{\system{}: Towards A Tag Model for Cross-Query Reuse in Semantic Query Processing [Vision]}

\author{GuangXue Zhang}
\affiliation{%
  \institution{University of California, Irvine}
  \city{Irvine}\country{USA}
  }
\email{gxzhang7@uci.edu}

\author{Nada Lahjouji}
\affiliation{\institution{Blizzard Inc}
\city{Irvine}\country{USA}
}
\email{nlahjouji@blizzard.com}

\author{Yiming Lin}
\affiliation{\institution{University of California, Berkeley}
\city{Berkeley}\country{USA}
}
\email{yiminglin@berkeley.edu}

\author{Ashwin Colaco}
\affiliation{\institution{University of California, Irvine}
\city{Irvine}\country{USA}
}
\email{acolaco@uci.edu}

\author{Sharad Mehrotra}
\affiliation{\institution{University of California, Irvine}
\city{Irvine}\country{USA}
}
\email{sharad@ics.uci.edu}

\date{July 2026}

\begin{abstract}
Large language models make it possible to pose semantic 
questions over diverse, multi-modal data. 
Recent systems give such workloads the 
declarative feel of a database
through \emph{semantic operators}: relational operators whose 
conditions a model judges.  
Most of them now run on a relational engine, yet almost none keeps what the model
says past the query: every judgment is discarded on completion and paid for
again the next time a query needs it.
We present the tag model and tag operators, the right unit for semantic
processing, and \system{}, a prototype built as a middleware over DBMS.
Two mechanisms are central. First, the \emph{tag}: each semantic answer a model
produces is written back to the relation as data, which the database engine
filters, indexes, and stores as it would any other.
Second, the \emph{tag operator}: the unit into which common semantic operators
are translated, which fixes the granularity of reuse.
A model judges a tuple once, and later queries reuse that judgment as data:
through relationships between predicates, a stored judgment can accept a tuple for
a later predicate, eliminate one that cannot satisfy it, simplify its test, or order
the search of a query that stops early.
On the movie scenario of SemBench, our prototype sends the model $10$ to
$108\times$ fewer rows than iPDB, ThalamusDB, Palimpzest, and LOTUS at broadly
competitive quality, and three of its eight queries issue no model call at all.
More broadly, a database that keeps its judgments lets its schema grow from
the questions its users ask, so that each semantic answer is paid for once, stays
consistent across queries, and can be audited through its provenance.
\end{abstract}

\maketitle

\vspace{-2mm}

\vldbtopmatter

\section{Introduction}
\label{sec:intro}

Large language models make it possible to evaluate, at query time, a class of conditions
relational databases cannot support: predicates whose satisfaction is computed over the
meaning of unstructured content rather than typed values. Consider applications that
hold large collections of multi-modal data, such as Amazon, whose catalogue carries
product descriptions, customer reviews along with pictures, or Netflix, whose movies
carry synopses and trailers. Given the number of customers such platforms serve, the
questions asked on them cannot be anticipated. Users phrase questions in their own
terms, so the set of predicates a workload will eventually ask is unbounded and unknown
when the schema is designed. A customer may search Amazon for \emph{``a Christmas-themed
gift that suits my grandparents''}, or parents may look for \emph{``a comedy that adults
enjoy but is still suitable for young children''}. No column of either schema records
whether a product is Christmas-themed or whether a film is suitable for children.
Instead, a model such as an LLM analyzes the question on the fly and judges which
products or movies semantically satisfy it.

Although such questions rarely repeat literally, the judgments their predicates require
often overlap: gift and holiday themes return every holiday season, and family-suitable
films are asked for by other families as well. 
An LLM agent that answers each question in isolation cannot exploit this overlap. 
The agent retains nothing between queries, so
every semantic judgment is discarded on completion and paid for again the next time it
is needed, and none of the pruning and reordering a database optimizer would apply is
available to it.
\emph{Semantic operator systems} answer such questions with relational-style operators
(filter, map, join) whose condition is written in natural language and judged by an LLM.

We compare recent \emph{semantic operator systems} in Table~\ref{tab:related_work}
from four angles: planner-level optimization, 
operator-level optimization, whether the system is built upon a DBMS, 
and whether it caches model results. 
Among the systems listed in Table~\ref{tab:related_work}, only LOTUS~\cite{lotus} and Palimpzest~\cite{palimpzest} 
are built as standalone engines, while the others build their semantic operators upon a DBMS
(inside DuckDB~\cite{duckdb} or Snowflake~\cite{snowflake}, as UDFs, or as a layer over DuckDB).
The last column compares whether a system caches model results; iPDB~\cite{ipdb}, PLOP~\cite{plop}, and FlockMTL~\cite{flockmtl} do,
but only for duplicate tuples
within a query,
and a cross-query cache is provided by LOTUS and AIDB~\cite{aidb}, under the constraint 
of an identical prompt, which limits the use case.

\begin{table}[t]
  \centering
  \caption{Comparison of current representative semantic query processing systems.
  \emph{Planner}: plan-level optimization.
  \emph{SemOps}: semantic operator-level optimization.
  \emph{DB-based}: integration with a relational DBMS.
  \emph{Cache}: caching of semantic inference results.
  Circles denote full, partial, and no support.}
  \vspace{-3mm}
  \begin{tabular}{lcccc}
    \toprule
     & Planner & SemOps & DB-based & Cache \\
    \midrule
    LOTUS~\cite{lotus}
        & \Circle     & \CIRCLE     & \Circle     & \LEFTcircle    \\
    Palimpzest~\cite{palimpzest}
        & \CIRCLE     & \Circle     & \Circle     & \Circle     \\
    iPDB~\cite{ipdb}
        & \LEFTcircle & \LEFTcircle & \CIRCLE     & \LEFTcircle \\
    PLOP~\cite{plop}
        & \CIRCLE     & \Circle     & \LEFTcircle & \LEFTcircle \\
    ThalamusDB~\cite{thalamusdb}
        & \LEFTcircle & \LEFTcircle & \LEFTcircle & \Circle     \\
    FlockMTL~\cite{flockmtl}
        & \Circle     & \CIRCLE     & \LEFTcircle & \LEFTcircle \\
    Sema~\cite{sema}
        & \CIRCLE     & \LEFTcircle & \CIRCLE     & \Circle     \\
    Sema-SQL~\cite{semasql}
        & \CIRCLE     & \LEFTcircle & \LEFTcircle & \Circle     \\
    Cortex AISQL~\cite{cortexaisql}
        & \LEFTcircle & \CIRCLE     & \CIRCLE     & \Circle     \\
    \bottomrule
  \end{tabular}
  \label{tab:related_work}
  \vspace{-7mm}
\end{table}


We propose a new approach, built on two notions: \emph{tag operators} and \emph{tags}.
A \emph{tag operator} is a semantic operator, only a more basic one: 
it takes a single semantic condition and issues one model call
about a single tuple, or about a single pair of tuples.
A \emph{tag} holds the value a model returns, as well as other information 
such as a confidence score, the model name, and a timestamp, which help track when each
tag was produced and at what quality.
A tag is an ordinary column the DBMS stores, indexes, and filters like any other,
read and written through SQL. 
This granularity is what makes such reuse possible: an answer produced for a whole table
at once can be reused only by a query asking the same question of the same data, 
whereas an answer stored in a tag serves any later tag operator and any future query
that needs that tuple judged. 
None of the DB-based systems in Table~\ref{tab:related_work} keeps a judgment 
past the query; at most it is kept within one. 
Reuse of arbitrary semantic conditions requires a data model designed for it.
With the small set of tag operators that we define in Section~\ref{sec:tag-ops}, 
we are able to express the semantic operators commonly used in other semantic 
operator systems, by decomposing them into a composition of tag operators 
and ordinary relational operators (Section~\ref{subsec:decomposition}),
but at a granularity at which the returned answers can be stored and reused.

Reuse also reaches beyond the identical question: through relationships
between predicates, which an LLM judges from their meaning, the
judgments a tuple already carries can accept it for a later predicate, eliminate it,
or simplify its test; where they only make an answer more likely, they can still
order the search of a query that stops early (Section~\ref{sec:reuse}).

To study the feasibility of this approach we built \system{}, a prototype that extends
a DBMS (PostgreSQL, but the design is not limited to it) with our tag model. Further,
\system{} aims to make writing the operator program optional: the user asks a question
in natural language, and an LLM-based parser produces the plan of operators from it,
though a plan can also be submitted directly. Queries that share semantic conditions
then grow cheaper over the workload, since the tags written for earlier queries already
answer part of new ones. 
On SemBench's~\cite{sembench} movie scenario, \system{} judges $10$ to $108\times$
fewer rows and pairs than four baselines, at competitive quality.
The savings are the visible part of a larger shift. Every tag is a column no
designer anticipated, added because a user asked, so the database's schema grows
from its workload. Because each judgment is stored with its provenance, the same
question receives the same answer across queries, and every answer can be
audited or re-judged. A database used this way does not merely call a model: it
accumulates what the model has established about its data, shared by every
application that queries it.

This paper is thus organized as follows:
Section~\ref{sec:tagmodel} presents that data model, the tag, and the design of the
semantic system built on a database that supports it. 
Section~\ref{sec:tag-ops} argues
that semantic operators are too coarse to be the unit of reuse, defines the tag
operators as the appropriate unit, and shows how the existing operators translate onto
them. 
Section~\ref{sec:reuse} shows how the judgments a tuple already carries decide,
simplify, or order the evaluation of later predicates.
Section~\ref{sec:study} reports a feasibility study with \system{}, and
Section~\ref{sec:agenda} lays out the directions database research must pursue to make
semantic operators first-class citizens of the database.

\section{The Tag Model}
\label{sec:tagmodel}

Everything in this paper rests on one idea: 
the answer a model produces should be data.
A \emph{tag} is a column added to a relation, 
or a relation of its own, that stores the
answer a model returned for a predicate about a tuple or a pair of tuples. 
For example, ``find the calorie of food that is vegan'' needs a semantic predicate
\emph{is\_vegan} and a semantic map \emph{calories}, both requiring a model judgment.
A tuple carries a value (\texttt{TRUE} or \texttt{FALSE} for a Boolean predicate such as
\emph{is\_vegan}, a typed value for a mapping such as \emph{calories}) if it has been
evaluated, and \texttt{NULL} if it has not. 
We call a tuple whose tag holds no value \emph{open tuple}.
When no tag exists yet for a predicate, one
is created on first use with a value that is \texttt{NULL} for every tuple, and filled
as evaluation proceeds. Tags are named for what they store (\texttt{tag\_is\_vegan},
\texttt{tag\_map\_calories}), keyed by the tuple identifier(s) the answer is about, and
read and written through ordinary SQL. 
Because a tag is an ordinary column, everything
the database does for data it now does for judgments: the engine filters, indexes, and
stores a tag as it would any other column, and once every tuple carries a value, a semantic
condition is answered by database with no model in the loop. 
The cost of a predicate
is therefore paid once for a tuple rather than once per query: as a workload
accumulates, more of each new query is answered by the database and less of it by a
model.

\noindent\textbf{Tag provenance.}
A tag stores more than the value; it stores the value's \emph{provenance}: the
confidence score the executing model assigned to it, the name of that model, the
execution timestamp, and, when semantic processing runs as a sequence 
of models of increasing capability (a cascade), the stage at which
the value settled. 
A tag value can be wrong because of the model's accuracy or hallucination
problems, and reusing it can turn a one-time error into a persistent one;
the confidence score lets a later query compare the stored value against its own quality
requirement and decide whether to trust it or evaluate the tuple again, so a
low-confidence answer is not silently inherited. 
For an ordinary model execution, the
score comes from the log probability the model reports for the token it emitted, which
costs nothing beyond the call itself. 
The model name and the stage together record how
much effort has already been spent on a tuple, so a query that is not satisfied with the
stored value resumes at the next stage rather than the first. The
timestamp records when that work happened, which lets the system find
the answers that were produced before a model was replaced or the data was updated. 
In our design a tag value is an ordinary column, or a relation of its own when the judgment 
is about a pair, and the provenance is kept in a side table keyed by the tag and the tuple or pair, 
so the engine stores, indexes, and filters them like any other column and table.

\noindent\textbf{Relation to other ways of storing model outputs.}
The ancestral idea is MauveDB's~\cite{mauvedb} model-based view, a relation whose
content a model computes; a tag can be read as a semantic view that is
\emph{partially} materialized, filled tuple by tuple as queries demand, where
MauveDB maintains its views as wholes. AIDB~\cite{aidb} materializes model outputs
sparsely and caches them in an RDBMS, but its reuse is by exact invocation, without
the provenance that lets a stored judgment be re-judged or answer a different
question. Enrichment databases such as EnrichDB~\cite{enrichdb} compute derived
attributes at query time and can materialize them, and semantic caches such as
GPTCache~\cite{gptcache} reuse a stored response when a new prompt is
embedding-similar to an old one; tags sit between these, materialized on demand by
the predicate the question itself introduces. None of these makes the per-judgment
unit a persistent, queryable part of the database.

\noindent\textbf{Tagging inside the database.} A database that supports tagging suffices for a semantic processing system 
realized as a thin layer over the DBMS. A question posed in
natural language, or as a program of semantic operators, is turned into a plan 
that separates the conditions SQL can express, which run first,
from those that require a model's judgment.
Each semantic condition routes through an operator that consults the relevant 
tag before any model is called: tuples the tag
already covers are answered by a database read, 
only the open tuples reach a model, and
every answer the model returns is written back to the tag. 
The \emph{tag operators} that behave this way, are the
subject of Section~\ref{sec:tag-ops}, and Section~\ref{sec:study} describes \system{}, the
prototype we built.

\section{Tag Operators: The Right Unit of Reuse}
\label{sec:tag-ops}

To give semantic processing declarative form, systems such as LOTUS~\cite{lotus},
Palimpzest~\cite{palimpzest}, and iPDB~\cite{ipdb} define \emph{semantic operators} as
primitives for expressing complex semantic queries: \texttt{sem\_filter},
\texttt{sem\_map}, \texttt{sem\_agg}, \texttt{sem\_groupby}, \texttt{sem\_join}, and
\texttt{sem\_topk} (Table~\ref{tab:decomposition}). Each carries a natural-language
argument that a model judges: a predicate, an extraction target, or a comparison
criterion.

Semantic operators are convenient to write, but they are too coarse to be the unit a
system reuses. We use the \texttt{sem\_groupby} operator as an example. 
Given a set of customer reviews about a product, 
a user asks for the five complaint topics over the reviews. 
Answering this complex query in LOTUS, for example, 
requires executing the \texttt{sem\_groupby} operator over the table of 
reviews, which involves multiple intermediate steps: 
first identifies the complaint label of each review and then groups the labels into five complaint topics.
However, if the system relies on one semantic operator \texttt{sem\_groupby}, what it produces is a single answer 
for this query over the given data, and the \emph{reusability} of that answer is limited to a 
query that asks exactly the same question over the same data.
Suppose another query instead asks for ten example reviews complaining about ``shipping issues'', 
one of the five complaint topics revealed by the previous query. 
It shares none of the work, although both need 
exactly the per-review judgment on the complaint label that \texttt{sem\_groupby} already paid for at its first step.

The first advantage of decomposing a complex operator is therefore a large gain in
workload reusability.
Once \texttt{sem\_groupby} is decomposed so that a map operator first extracts
the complaint label of each review (Section~\ref{subsec:decomposition}), storing each
label as a tag
makes it reusable by \emph{any} later query whose decomposition contains
the same map --- not only by another \texttt{sem\_groupby}.

The second advantage is flexibility in where an optimization applies: any algorithm
designed for the map operator can be embedded into the physical implementation of
\texttt{sem\_groupby}, cleanly separating the logical definition from the physical
implementation. An optimization such as a cascade~\cite{automix, frugalgpt}, which is defined for a single per-tuple judgment, can be attached to the
map inside \texttt{sem\_groupby} without changing what \texttt{sem\_groupby} means.

The principle the rest of the paper builds on follows: \emph{the unit of reuse and
optimization is not the semantic operator but the single model judgment} --- one
predicate applied to one tuple, or to one pair of tuples. Persist that unit, and reuse
crosses operator boundaries: the same per-review label serves a groupby today and a
filter tomorrow. Attach optimizations at that unit, and every operator built on it
benefits at once.
It is also the unit at which a stored judgment can settle a later predicate by itself, or narrow what the model must still judge. (Section~\ref{sec:reuse}).

\subsection{The Tag Operators}
\label{subsec:tag-ops}

A tag operator realizes exactly this unit. Most tag operators issue one or more calls to
a model (an LLM for filter, map, and compare, or an embedding model for clustering)
over a table $T$ and persist the result as a tag, so that a later query demanding the
same semantic condition can read the answer with ordinary SQL instead of calling the
model again. Let $P$ denote the natural-language argument passed to the model, e.g., a
predicate such as \emph{is\_vegan} for a filter or an extraction target such as
\emph{calories} for a map. There are five tag operators\newstuff{.}
\texttt{octo\_filter} and \texttt{octo\_map} are identical to
\texttt{sem\_filter} and \texttt{sem\_map}: each issues a single model call per tuple
$t \in T$, so they are already the most fine-grained operators and need no further
decomposition.
\texttt{octo\_compare} issues one model call per pair $(t,s)$;
because no single tuple's row can hold an answer about two tuples, its tag is not a
column but a relation, with one row per pair judged, keyed by $(t,s)$.
\texttt{octo\_clustering} groups the tuples of $T$ into $k$ clusters using an
embedding model rather than an LLM call, e.g., $k$-means over the values that
\texttt{octo\_map} extracted for predicate $P$.
\texttt{nest} collapses a relation into fewer rows by grouping,
or into a single row, as when a list of documents is summarized; it
issues no model call and stores no tag, and is realized
directly by \texttt{ARRAY\_AGG}.

\vspace{-2mm}
\subsection{Translating Semantic Operators}
\label{subsec:decomposition}

Each semantic operator decomposes into a composition of tag operators and ordinary
relational operators (\texttt{sql\_filter}, \texttt{sql\_join}, \texttt{sql\_groupby},
\texttt{sql\_agg}, \texttt{sql\_orderby}, \texttt{sql\_limit}).
Table~\ref{tab:decomposition} lists the translation for each operator; the translation
itself is fixed and adds no model call, so the result is correct given the
decomposition and separability judgments that produced the plan, whose validation
Section~\ref{sec:agenda} lists as open.
We work through one translation here; the rest follow a similar strategy.

Consider again the five complaint topics from the beginning of this section:
\texttt{sem\_groupby} with categories not given, so that a label has to be produced for each review before any group exists, becomes
$\texttt{sql\_agg} \circ \texttt{sql\_groupby} \circ \texttt{octo\_clustering} \circ
\texttt{octo\_map}$: an \texttt{octo\_map} extracts a complaint label per review and
stores it as a tag, \texttt{octo\_\allowbreak clustering} groups the labels into $k$ clusters, and
grouping and counting are ordinary SQL. Every intermediate judgment survives as a tag
that any later query can read.

One judgment the translation turns on deserves note: a \texttt{sem\_join} is
\emph{separable} when its condition can be decided by extracting a property from each tuple and comparing the two, and
\emph{non-separable} otherwise. iPDB applies the same decomposition, splitting a join that asks whether a CPU
and a motherboard are compatible into an extraction of the socket type from each side and
an ordinary match on it; in our terms an \texttt{octo\_map} followed by a
\texttt{sql\_join}, so the cost falls from $|R| \times |S|$ model calls to $|R| + |S|$. 
When the join is non-separable, every pair is judged directly
by \texttt{octo\_compare}, and the quadratic cost is inherent. \texttt{sem\_topk} splits
the same way. This distinction is where the largest savings of
Section~\ref{subsec:evidence} come from. Across every translation in
Table~\ref{tab:decomposition}, a model is called only inside a tag operator; every
other step is work the database already does.

\begin{table}[tb]
  \centering
  \footnotesize
  \setlength{\tabcolsep}{5pt}
  \caption{Semantic operators and their translation to tag operators.}\label{tab:decomposition}
  \vspace{-3mm}
  \begin{tabular}{@{}>{\raggedright\arraybackslash}p{0.24\columnwidth} >{\raggedright\arraybackslash}p{0.68\columnwidth}@{}}
    \toprule
    Sem Ops & Translation \\
    \midrule
    \texttt{sem\_filter}  & \texttt{octo\_filter} \\
    \texttt{sem\_map}     & \texttt{octo\_map} \\
    \texttt{sem\_agg}     & \emph{attribute already a column:} \texttt{sql\_agg} $\circ$ \texttt{octo\_filter} \\
                          & \emph{attribute not yet a column:} \texttt{sql\_agg} $\circ$ \texttt{octo\_map} $\circ$ \texttt{octo\_filter} \\
    \texttt{sem\_groupby} & \emph{categories given:} \texttt{sql\_agg} $\circ$ \texttt{sql\_groupby} $\circ$ \texttt{octo\_map} \\
                          & \emph{categories not given:} \texttt{sql\_agg} $\circ$ \texttt{sql\_groupby} $\circ$ \texttt{octo\_clustering} $\circ$ \texttt{octo\_map} \\
    \texttt{sem\_join}    & \emph{separable:} \texttt{sql\_join} $\circ$ \texttt{octo\_map} \\
                          & \emph{non-separable:} \texttt{sql\_filter} $\circ$ \texttt{octo\_compare} \\
    \texttt{sem\_topk}    & \emph{separable:} \texttt{sql\_limit} $\circ$ \texttt{sql\_orderby} $\circ$ \texttt{octo\_map} \\
                          & \emph{non-separable:} \texttt{sql\_limit} $\circ$ \texttt{sql\_orderby} $\circ$ \texttt{sql\_agg} $\circ$ \texttt{octo\_compare} \\
    \bottomrule
  \end{tabular}
  \vspace{-5mm}
\end{table}

\begin{figure*}[t]
\centering
\includegraphics[width=0.9\textwidth]{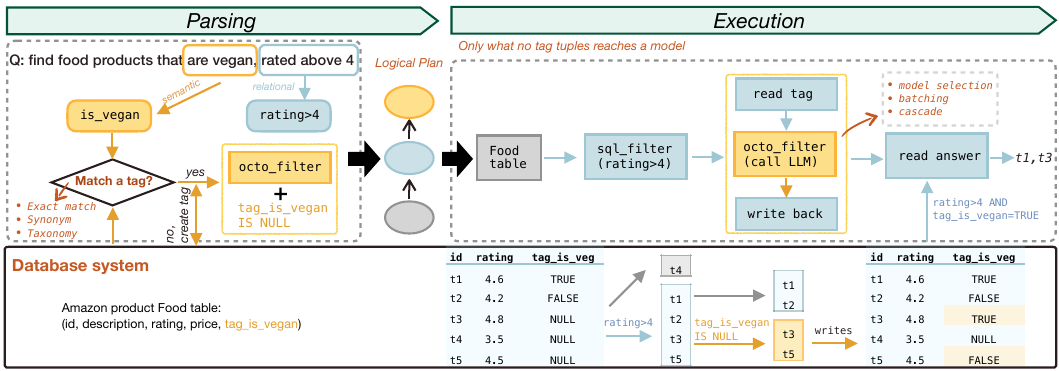}
\vspace{-3mm}
\caption{\system{} prototype on one query example. }
\label{fig:walkthrough}
\vspace{-3mm}
\end{figure*}

\section{Reusing Judgments}
\label{sec:reuse}

A tag serves wider than a cache. Reusing a tag is not limited to the identical
semantic condition: the tags of a tuple, one per judgment (Section~\ref{sec:tag-ops}),
record what a model has established about it, and a later request can often
be decided, or made cheaper, from them.

\noindent\textbf{Tags as known facts.}
Consider a tuple $r$ on which tags $t_1,\dots,t_n$ have been evaluated. What is known
about $r$ is the conjunction of existing tags
$V(r) = (t_1 = v_1) \wedge \dots \wedge (t_n = v_n)$, where
$v_i$ is \texttt{TRUE} or \texttt{FALSE} for a binary tag such as
\texttt{tag\_is\_vegan}, and a typed value for a map such as
\texttt{tag\_map\_calories}; a tag not yet evaluated on $r$ holds \texttt{NULL} and
contributes nothing. 
A new request asks whether $r$ satisfies a condition $Q$  on a tag $t$: that it
equals a value, or falls in a set or range of values.
Tags relate to one another
through \emph{rules} between their predicates, such as
$(\textit{is\_vegan} = \texttt{TRUE}) \rightarrow 
(\textit{is\_vegetarian} = \texttt{TRUE})$;
we write $\mathcal{O}$ for the rules known to the system. 
In the current prototype, the Parser's
LLM derives them from common knowledge when a query arrives; 
a taxonomy, knowledge
graph, or ontology could supply them instead (Section~\ref{sec:agenda}). 
A rule is derived once for
a pair of predicates, not once for a tuple.

\noindent\textbf{Reuse as entailment.}
A tuple $r$ is \emph{accepted} with no model call  
if the rules $\mathcal{O}$ and what is
already known about $r$, that is, $V(r)$, entail the condition $Q$; 
\emph{eliminated} if they entail its negation; 
and \emph{open} otherwise.
Table~\ref{tab:reuse} lists the four rules through which a single known fact
$t_i = v_i$ of $r$ bears on $Q$. 
Take rule (3), where the fact is necessary for $Q$:
\emph{vegan} requires \emph{vegetarian}, so a tuple with
\texttt{tag\_is\_vegetarian = FALSE} is eliminated from a query for vegan food; 
a tuple with \texttt{tag\_is\_vegetarian = TRUE} is not decided, 
but its test simplifies, 
since vegan food is vegetarian food that excludes dairy, eggs, and honey, which
we denote by the residual condition $Q'$.
Therefore, $Q$ is equivalent on
$r$ to $(t_i = v_i) \wedge Q'$, and only the residual $Q'$ needs testing,
by a cheaper mechanism such as a keyword match over the ingredient list. 
If the residual has already been answered, the conjunction of the tags in $Q'$
can decide what none decides alone:
the facts of a vegetarian tuple $r$ that also carries
\texttt{tag\_hasDairy\_or\_hasEggs = FALSE} and
\texttt{tag\_hasHone = FALSE} entail \emph{vegan}.
A map tag goes further: \texttt{tag\_map\_calories = 200} answers any request that
compares calories with a bound, by range containment, which SQL decides without
rules or a model. Storing values therefore serves more
requests than storing verdicts.
The other rules read the same way, and exact match, synonym, complement, and taxonomy are
their instances. 
Every rule compiles into the \texttt{WHERE} clause through a query rewriting approach~\cite{queryanswering},
that selects the open tuples, so reuse adds no model call per tuple, and as long as the data and $\mathcal{O}$ do
not change, $V(r)$ only grows, 
and with it the set of tuples decided without a model. 

\begin{table}[tb]
  \centering
  \footnotesize
  \setlength{\tabcolsep}{3pt}
  \caption{How a known fact $t_i = v_i$ of tuple $r$ decides a request $Q$.}
  \label{tab:reuse}
  \vspace{-3mm}
  \begin{tabular}{@{}>{\raggedright\arraybackslash}p{0.22\columnwidth}
                  >{\raggedright\arraybackslash}p{0.1\columnwidth}
                  >{\raggedright\arraybackslash}p{0.19\columnwidth}
                  >{\raggedright\arraybackslash}p{0.42\columnwidth}@{}}
    \toprule
    Rule & $r$ has & Outcome & Example ($Q$ \,/\, known fact) \\
    \midrule
    (1) $(t_i{=}v_i) \rightarrow Q$ & $t_i = v_i$ & accept & \emph{vegetarian} / vegan${=}$T (taxonomy); exact match, synonym \\
    (2) $(t_i{=}v_i) \rightarrow \neg Q$ & $t_i = v_i$ & eliminate & \emph{meat lover} / vegan${=}$T (complement, \texttt{NOT}) \\
    (3) $Q \rightarrow (t_i{=}v_i)$ & $t_i \neq v_i$ & eliminate & \emph{vegan} / vegetarian${=}$F \\
    & $t_i = v_i$ & test residual $Q'$ & \emph{vegan} / vegetarian${=}$T \\
    (4) $Q \rightarrow \neg(t_i{=}v_i)$ & $t_i = v_i$ & eliminate & contrapositive of (2) \\
    none entailed & & open & judged by a model, ordered by $P(Q \mid V(r))$ \\
    \bottomrule
  \end{tabular}
  \vspace{-4mm}
\end{table}

\noindent\textbf{Probabilistic reuse.}
Entailment is the limit of a probabilistic view. 
Where $\mathcal{O}$ and $V(r)$ entail neither $Q$ nor $\neg Q$, 
$V(r)$ still changes how likely $Q$ is.
Acceptance and elimination are the cases where $P(Q \mid V(r))$
is $1$ and $0$.
A rule may hold only with probability $P(Q \mid t_i = v_i)$. 
For example, $P(\textit{vegan}\mid\texttt{tag\_hasChocolate=TRUE})$ 
is not necessarily $0$, but close to it, since few chocolate foods are vegan.
Between $0$ and $1$ the probability can
serve as a decision, accepting or eliminating $r$ once it is close enough to $1$ or
$0$, which is safe only under a quality guarantee~\cite{kang2020supg}; 
or as an order, which is always safe: for a query that stops early, with a \texttt{LIMIT} or \texttt{EXISTS}, 
any tuples that satisfy $Q$ make a correct answer, 
so judging the open tuples in
decreasing $P(Q \mid V(r))$ reaches the stopping condition 
with fewer model calls.
The tag
store can also estimate the rules themselves: tuples on which both $t_i$ and $Q$ have
been judged estimate $P(Q \mid t_i = v_i)$, 
so it accumulates statistics on how predicates correlate,
much as a database keeps histograms over its columns.

\noindent\textbf{Related ideas.}
Deciding a request from known facts and rules aligns with two lines: semantic query
optimization~\cite{king1981quist, chakravarthy1990sqo} decides a predicate from
declared constraints, and answering queries using views or semantic caching~\cite{queryanswering, chaudhuri1995optimizing, dar1996semantic} splits a query into
what is already stored and what remains.
Both start from a schema and a rule set fixed at design time.
The tag model lets them grow as queries arrive, 
a predicate the schema never carried becoming a column and the rules
relating predicates being derived on demand, at the price that our rules are beliefs
that model judgments may violate, not constraints.

\section{A Feasibility Study: \system}
\label{sec:study}

\subsection{The \system{} Prototype}
\label{subsec:prototype}

Figure~\ref{fig:walkthrough} depicts how a running
example \emph{``find food products that are vegan, rated above
4''} over Amazon product is handled in \system{}.

\noindent\textbf{Parsing.}
An LLM-based agent \emph{Parser} first receives the query and analyzes 
the predicates while reading the existing schema in the database. 
It separates the relational predicate \texttt{rating > 4} from the semantic
predicate \emph{is\_vegan}, and checks which stored tags are related to the latter,
and how (Section~\ref{sec:reuse}).
If no tag is reusable, the Parser creates the tag column 
(\texttt{tag\_is\_vegan}) in the relation, 
whose value is \texttt{NULL} for every tuple.
It then outputs a logical plan with
the relational filter and \texttt{octo\_filter},
translated from \texttt{sem\_filter} (Table~\ref{tab:decomposition}).
The current prototype accepts and eliminates tuples through exact match, synonym,
complement, and taxonomy; simplifying the test and ordering the search are not yet implemented.

Parsing makes exactly one call to
a state-of-the-art LLM per query, a fixed cost independent of the size of the
database.
The logical plan pushes down all relational conditions to ensure they execute first
before any model judgment; 
since the natural-language interface rests on this one component,
the Parser is constrained to emit a machine-validated typed schema.
The logical plan then passes to the \emph{Executor}, which decides
how it executes,
a question of cost rather than of meaning, and so needs no
model of its own: it hands every relational operator to the DBMS and decides,
for each tag operator, how the calls no tag settles are issued.

\noindent\textbf{Execution.}
Execution proceeds in three steps, which Figure~\ref{fig:walkthrough}
continues. 
\emph{Read first}: 
the database applies \texttt{rating > 4} and prunes the tuples it excludes,
and reads the value of \texttt{tag\_is\_vegan} for each tuple in one SQL query.
If the user sets a threshold 
$\theta$ on the tag value's confidence score,
the tag operator will select \emph{open} tuples by: 
\texttt{tag\_is\_vegan IS NULL OR conf\_is\_vegan < $\theta$}.
The confidence score joins in from the tag provenance
table of Section~\ref{sec:tagmodel} and its threshold defaults to $0$,
trusting every stored value as it stands. 
A related tag enters the same query: when the Parser has found, say, that
\texttt{tag\_is\_vegetarian = FALSE} eliminates a tuple for \emph{is\_vegan}
(Section~\ref{sec:reuse}), that condition joins the \texttt{WHERE} clause, and such
tuples never become open.
\emph{Settle}: 
the model judges the open tuples alone, one tuple per judgment (or one pair when
the operator is \texttt{octo\_compare});
the prototype issues the calls
concurrently and batches several tuples into one prompt, and the Executor
could add more optimization strategies such as 
model selection~\cite{octoselector} and cascade execution~\cite{frugalgpt,lotus,lazystack}. 
\emph{Write back}: 
each answer persists in its tag with its provenance (Section~\ref{sec:tagmodel}). Once
the tuples the query needs carry a value (Section~\ref{subsec:evidence}), 
the query is answered by the database alone, 
as \texttt{rating > 4 AND tag\_is\_vegan = TRUE}.
The judgments this query paid for are data for whatever query
comes next. 
Execution
may be iterative rather than a single pass, stopping when its own conditions are met.

\subsection{Case Studies}
\label{subsec:evidence}

\begin{table}[t]
\centering
\footnotesize
\setlength{\tabcolsep}{5pt}
\begin{tabular}{@{}l rrrrr@{}}
\hline
\textbf{Query} & \textbf{\system{}} & \textbf{iPDB} & \textbf{ThalamusDB} & \textbf{PZ} & \textbf{LOTUS} \\
\hline
Q1 (filter, \texttt{LIMIT}) & 20  & \textbf{10} & 20 & 15 & 1,865 \\
Q2 (filter, \texttt{LIMIT}) & \textbf{44}  & 119 & 60 & 79 & 120 \\
Q3 (count) & \textbf{76}  & 119 & 119 & 120 & 120 \\
Q4 (ratio) & \textbf{0}   & 119 & 119 & 120 & 120 \\
Q5 (join, \texttt{LIMIT}) & \textbf{20}  & 140 & 200 & 100 & 8,209 \\
Q6 (join, \texttt{LIMIT}) & \textbf{0}   & 140 & 100 & 100 & 9,809 \\
Q7 (join) & \textbf{108} & 257 & 16,384 & 16,384 & 8,547 \\
Q8 (group by) & \textbf{0}   & 1,864 & 238 & 120 & 120 \\
\hline
\textbf{\#rows/pairs judged} & \textbf{268} & 2,768 & 17,240 & 17,038 & 28,910 \\
Mean quality & 0.82 & 0.87 & 0.83 & 0.79 & 0.79 \\
\hline
\end{tabular}
\caption{Cost in terms of numbers of rows judged by the model per query on SemBench movie (Q1--Q8).}
\label{tab:reuse_movie}
\vspace{-10mm}
\end{table}

\begin{figure}[t]
\centering
\includegraphics[width=\columnwidth]{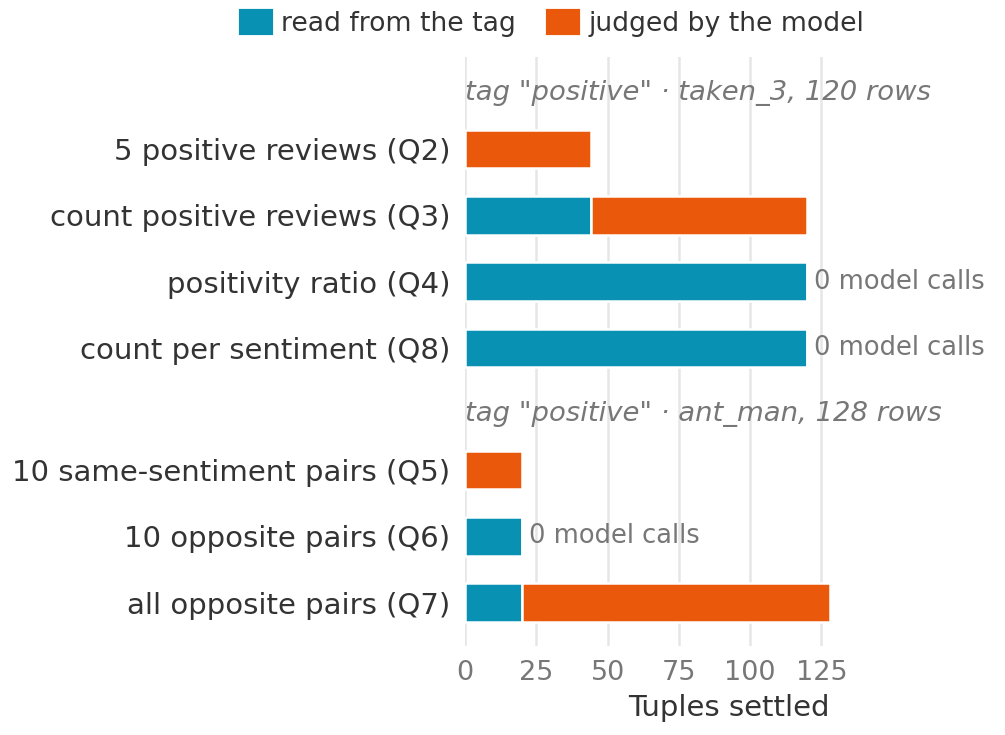}
\vspace{-6mm}
\caption{The two groups of queries that share a tag: per query, the settled
tuples read from the tag (teal) or judged by the model (orange).}
\label{fig:reuse}
\vspace{-6mm}
\end{figure}

Three case studies on SemBench's~\cite{sembench} movie scenario, chosen
  because its data is real (Rotten Tomatoes reviews) and its predicates recur
  across queries, examine \system{}'s and each baseline's capacity for reuse,
  intra-query and inter-query, plus the savings due to the database
  itself, the relational guard and the separable join.
  We compare \system{} with iPDB~\cite{ipdb},
  which extends DuckDB in-engine, ThalamusDB~\cite{thalamusdb}, which
  runs on DuckDB, LOTUS~\cite{lotus} and Palimpzest~\cite{palimpzest}, which run
  their own engines.
  All systems run under one setup:
  \texttt{gemini-2.5-flash}, $20$ concurrent calls, with $1,865$ \texttt{Reviews} and $116$
  \texttt{Movies} with duplicates.

The measure on cost is independent of pricing and of batching: 
cost is counted by the number of rows (or pairs, for a join) the model judges. Batch size is set to one where a system exposes it; where a join packs several pairs into one prompt (iPDB, ThalamusDB), we count the pairs the prompt carries. 

\noindent\textbf{Case study 1: Reuse within a query.}
Within a query, the same input can reach the model more than once: some duplicates
are in the data (the benchmark's Movies table lists 11 of its 105 films twice), and
some are made by the plan, when a join fans one tuple out into many rows before a
semantic filter reads it. \system{} writes each model result to a tag on the base
relation, one cell per tuple and predicate, which removes both, as the temporary
per-query cache tables of iPDB and PLOP do. For \emph{``find family-friendly movies
reviewed by a top critic''} under a pull-up plan, Palimpzest, with no cache, passes
448 joined rows to the model; a cache keyed on the operator's input, as in iPDB or
through \system{}'s per-movie tag, collapses them to 81, the number of distinct movies.

\noindent\textbf{Case study 2: Reuse across queries.}
None of the systems we run keeps a judgment past the query: iPDB drops its table
after \texttt{PREDICT} returns, Palimpzest keeps none, and LOTUS's optional cache is
keyed on the rendered prompt, so a later query reuses an answer only if it renders the
same text for the same row.
Over Q1--Q8 (Table~\ref{tab:reuse_movie})
\system{} judges 268 rows against 2,768 for iPDB, 17,240 for ThalamusDB, 17,038 for
Palimpzest and 28,910 for LOTUS, that is $10\times$, $64\times$, $64\times$ and
$108\times$ fewer. The gap to iPDB is the reuse across queries alone, since both
delegate the relational work to a DBMS and both reuse within a query; the gap to the
other three adds the reuse within a query of case study 1 and the separable join of
case study 3.

Figure~\ref{fig:reuse} traces the queries that share a tag: within a group the first
query pays for the tag, and later queries read what it holds, under the rules of
Table~\ref{tab:reuse}. By rule (1), exact match: Q2 and Q3 fill
\texttt{tag\_positive} for \emph{taken\_3} and Q5 fills it for \emph{ant\_man}, so Q4,
Q8 and Q6 issue no model call, milliseconds of pure SQL. By rule (2), complement: Q8
and Q6 ask for opposite-sentiment reviews, answered with \texttt{NOT} on
\texttt{tag\_positive}. By rule (1) again, taxonomy: Q1's \emph{clearly positive} is
narrower than Q2's \emph{positive}, so a tuple Q1 answered \texttt{TRUE} is settled
for Q2, while a \texttt{FALSE} one stays open. Reuse is also partial: Q3 reuses the 44
tuples Q2's \texttt{LIMIT} settled, and Q7 starts from the 20 that Q5 settled.

Reuse has a price in quality: Q6 reads the tags Q5 left behind and scores 0.58, while
Q7 judges the rest of the 128 reviews and reaches 0.74, so an inheriting query may
repeat whatever earlier judgments got wrong. The confidence threshold $\theta$
(Section~\ref{subsec:prototype}) and the strictness with which the Parser matches a
new predicate to stored tags are the two knobs; Section~\ref{sec:agenda} takes up
both. SemBench runs each query once, in a fixed order, the least favorable case for
reuse; in a real workload, where users repeat or rephrase
queries~\cite{esci,workloadweaver}, the savings would only grow.

\noindent\textbf{Case study 3: relational guard and separable join.}
These savings come with the database. \emph{Relational guard}: a relational filter,
such as the movie name in every query except Q1, is applied before any model is
consulted, and a \texttt{LIMIT} stops evaluation early: on Q1, Q5 and Q6, \system{}
and iPDB stop once the SQL query meets it, Palimpzest pushes down a logical
\texttt{LIMIT}, and LOTUS judges all 1,865. \texttt{DISTINCT}, \texttt{HAVING} and
\texttt{EXISTS} settle early the same way, though SemBench does not exercise them.
\emph{Separable join}: Q5--Q7 ask for pairs of \emph{ant\_man} reviews with the same
or opposite sentiment, a property of one review, so a tag operator judges each of the
128 once and a SQL self-join pairs the verdicts. On Q7, Palimpzest and ThalamusDB
judge all 16,384 pairs, LOTUS 8,547, and iPDB 257 rows (it decomposes the join but
judges each side separately), against 108 for \system{}, at quality 0.74, the best of
the five (0.24--0.73 for the others).

\vspace{-4mm}
\section{Future Directions}
\label{sec:agenda}


\system{} is a prototype, but the position it argues points further: a
database whose semantic schema is built by its users' questions, whose judgments are
shared by every application that queries it, and whose answers stay consistent
as the models beneath it change. 

{\em When to trust a reused answer.}
A stored judgment may itself be wrong (Section~\ref{sec:tagmodel}), so the Executor
re-judges any value whose confidence falls below a threshold, which could give a
query a \emph{quality guarantee}. Reuse through a rule (Section~\ref{sec:reuse})
adds a second source of error: a rule is inferred rather than declared, and a wrong
rule affects the queries that it judges. Whether a model-reported score can hold a
guarantee, how much evidence a rule needs before it is trusted, and how to combine
per-rule probabilities into an order over open tuples, are open.

{\em The tag lifecycle.}
A tag is derived data, and the store that holds it needs maintenance, most of it offline:
tags a workload rarely reads can be evicted to reclaim space;
tags whose tuples have changed must be refreshed;
and tags holding \texttt{NULL} can be filled while the system is idle.
This is the view-maintenance problem~\cite{mauvedb} in a new
setting: a tag is a partially materialized semantic view, and it needs maintaining
not only when the data changes but when the model or the prompt that defines it does.
Provenance records when a value was produced, but semantic staleness has no full
treatment yet.

{\em The parser as the trusted base.}
  The Parser 
  decides the order of the tag operators,
  the decomposition, and which stored tags a predicate may reuse.
  All three now rest on a frontier LLM
  guided only by a few instructions.
  Where one judgment falls short, the literature offers grounding in external
  knowledge, such as deriving the relationships between predicates from a taxonomy
  or knowledge graph rather than from the LLM alone~\cite{chen2023contextual}, agreement across an ensemble read as a
  confidence~\cite{Self-consistency}, and abstention when that confidence is
  low~\cite{abstention}.
  How accurate the reuse judgment must be before permissive reuse is safe, and how
  that accuracy would be measured, are unstudied.

{\em Into the engine.}
A cost-based optimizer that prices model calls next to I/O could reorder semantic
and relational work rather than always pushing SQL first, and write-back during reads
raises transactional questions the middleware currently sidesteps. 


\bibliographystyle{ACM-Reference-Format}
\bibliography{s9_bib}

\end{document}